\documentstyle[prl,aps,psfig]{revtex}

\draft

\def\fun#1#2{\lower3.6pt\vbox{\baselineskip0pt\lineskip.9pt
  \ialign{$\mathsurround=0pt#1\hfil##\hfil$\crcr#2\crcr\sim\crcr}}}
\newcommand{\beq}{\begin{equation}}
\newcommand{\eeq}{\end{equation}}
\begin{document}
\twocolumn[\hsize\textwidth\columnwidth\hsize\csname
@twocolumnfalse\endcsname


\draft
\title{On the Origin of the Highest Energy Cosmic Rays}
   
\author{F. W. Stecker}

\address{Laboratory for High Energy Astrophysics, Code 661,
NASA/Goddard Space Flight Center, Greenbelt, MD 20771, USA.}

\vspace{0.8cm}

\maketitle

\begin{abstract}

We present the results of a new estimation of the photodisintegration and
propagation of ultrahigh energy cosmic-ray (UHCR) nuclei in intergalactic
space. The critical interactions for energy loss and photodisintegration
of UHCR nuclei occur with photons of the infrared background radiation (IBR). 
We have reexamined this problem making use of a new determination of the IBR
based on empirical data, primarily from IRAS galaxies, and also collateral 
information from TeV $\gamma$-ray observations of two nearby BL Lac objects.
Our results indicate that a 200 EeV Fe nucleus can propagate $\sim$ 100 Mpc
though the IBR. We argue that it is possible that the highest energy cosmic
rays observed may be heavy nuclei.

\end{abstract}

\vskip 2pc]

Shortly after the discovery of the cosmic microwave background radiation (CBR),
it was shown that cosmic rays above $\sim$100 EeV ($10^{20}$eV) should be 
attenuated by photomeson interactions with CBR photons ~\cite{gzk}. It was later 
calculated that heavier nuclei with similar {\it total} energies would also be
attenuated, but by a different process, {\it viz.}, photodisintegration 
interactions with IBR photons ~\cite{psb}. The particular detections of two events with
energies well above these expected cutoffs, one at $\sim 200$ EeV ~\cite{hayash} and one
at $\sim 300$ EeV ~\cite{bird} have provided a double problem for cosmic-ray physicists.
How does nature accelerate particles to these extreme energies and how do they
get here from extragalactic sources ~\cite{elsom}? To answer these questions, new 
physics has been invoked, physics involving the formation and annihilation of 
topological defects (TDs) which may have been produced in the very earliest 
stages of the big bang, perhaps as a result of grand unified theories. TD
annihilation has unique observational consequences, such as the copious 
production of UHCR neutrinos and $\gamma$-rays ~\cite{sigl} and an interesting
satellite experiment called {\it OWL} has been proposed to test look for such
consequences ~\cite{ormes}. However, we will reexamine here the more conventional
scenario by which UHCRs are accelerated at extragalactic sites.

Although cosmic acceleration to energies above 100 EeV pushes our present 
theoretical ideas to their extreme, it has been suggested that such 
acceleration may occur in hot spots in the lobes of radio galaxies ~\cite{bier}. For
the purposes of this {\it Letter}, we will assume that such acceleration
processes can occur in nature. We now turn specifically to the propagation 
problem. A UHCR proton of energy $\sim$ 200 EeV has a lifetime against 
photomeson losses of $\sim 3\times 10^{15}$s; one of energy 300 EeV has a
lifetime of about half that ~\cite{stecker}. These values correspond to linear propagation
distances of $\sim$ 30 and 15 Mpc respectively. Even shorter lifetimes were
calculated for Fe nuclei, based on photodisintegration off the IBR ~\cite{psb}.
Recent estimates of the lifetimes of UHCR $\gamma$-rays against 
electron-positron pair production interactions with background radio photons
give values below $10^{15}$s ~\cite{proth}. Within such distances, it is difficult to
find candidate sources for UHCRs of such energies.

In this {\it Letter}, we present the results of a new estimation of the 
photodisintegration and propagation of UHCR nuclei through the IBR in 
intergalactic space.
We have reexamined this problem making use of a new determination of the IBR
based on empirical data, primarily from IRAS galaxies, recently calculated by
Malkan and Stecker ~\cite{ms}. Malkan and Stecker calculated the intensity and 
spectral energy distribution (SED) of the IBR based on empirical data, some 
of which was obtained for almost 3000 IRAS galaxies. It is these sources 
which produce the IBR. The data used for this calculation included (1) the 
luminosity dependent SEDs of these galaxies, (2) the 60 $\mu$m luminosity
function for these galaxies, and (3) the redshift distribution of these
galaxies. The magnitude of the IBR flux derived by these authors
is approximately an order of magnitude lower that that used by Puget, Stecker
and Bredekamp ~\cite{psb} in their extensive examination of photodisintegration of UHCR
nuclei. This determination of a lower value for the magnitude of the IBR is
also indicated by the observed lack of strong absorption in the multi-TeV
$\gamma$-ray spectra of the active galaxies known as the BL Lac objects Mrk 421
~\cite{whip} and Mrk 501 ~\cite{ahar}. The lack of an obvious absorption feature up to an 
energy greater than $\sim$ 5-10 TeV is consistent with the new, lower value 
for the IBR ~\cite{sd}. 

The Malkan-Stecker SED of the IBR has a similar shape to the one labeled 
``HIR'' in the paper of Puget, {\it et al.} ~\cite{psb} in the mid-IR and far-IR 
range. However, it is approximately an order of magnitude lower in intensity.
Therefore, one may replace the lifetimes given in ref. ~\cite{psb} by lifetimes
which are longer by a factor of 10-20. Here we conservatively assume a factor
of 10. Figure 1 is adapted from Fig. 14 of Puget, {\it et al} ~\cite{psb}. It 
indicates how a flux of UHCR Fe nuclei with an initial $E^{-3}$ differential
power-law spectrum will develop a cutoff at a critical energy, $E_{c}$ which
has an inverse dependence on the propagation time. In fact, for energies in
the range near 200 EeV, $E_{c} \simeq 150 (ct/100 Mpc)^{-1/2}$ EeV.

\begin{figure}
\centerline{\hskip 0.4cm\psfig{file=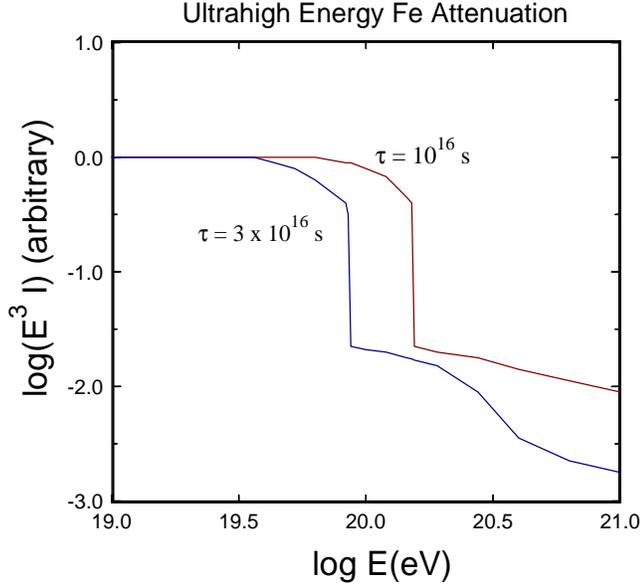,height=4.0in}}
\vskip -1.0cm
\caption[]{Attenuated $E^{-3}$ differential spectrum for ultrahigh energy Fe
nuclei for propagation times as indicated.}
\label{Figure 1}
\end{figure}

We conclude from this that the highest energy CR induced air-showers could
have been produced by UHCR nuclei propagating from a distance of the order
of 100 Mpc! Stanev, Biermann and Lloyd-Evans ~\cite{sbl} 
have examined the arrival
directions of the highest energy air-shower events. These authors have pointed
out that the 200 EeV event [3] is within 10$^\circ$ of the direction of the 
powerful radio galaxy NGC315 and the 300 EeV event is within 12$^\circ$ of the
powerful radio galaxy 3C134. NGC315 lies at a distance of only $\sim$ 60 Mpc 
from us. The distance to 3C134 is unfortunately unknown because its location
behind a dense molecular cloud in our Galaxy obscures the spectral lines 
required for a measurement of its redshift.

It should be also pointed out that it is reasonable to expect that the highest
energy cosmic rays may be nuclei. This is because the maximum energy to which
a particle can be accelerated in a source of a given size and magnetic field
strength is proportional to its charge, $Ze$. That charge is 26 times larger 
for Fe than it is for protons. We conclude that is indeed possible that the
highest energy cosmic rays which have been observed are heavy nuclei from
radio galaxies which lie at distances within 100 Mpc from Earth.

\end{document}